41# Total Ionizing Dose Effects on Threshold Switching in 1*T*-TaS$_2$ Charge Density Wave Devices

G. Liu, *Member, IEEE*, E. X. Zhang, *Senior Member, IEEE,* C. D. Liang, *Student Member, IEEE*, M. A. Bloodgood, T. T. Salguero, D. M. Fleetwood, *Fellow, IEEE,* and A. A. Balandin, *Fellow, IEEE**Abstract* — The 1*T* polytype of TaS$_2$ exhibits voltage-triggered threshold switching as a result of a phase transition from nearly commensurate to incommensurate charge density wave states. Threshold switching, persistent above room temperature, can be utilized in a variety of electronic devices, e.g., voltage controlled oscillators. We evaluated the total-ionizing-dose response of thin film 1*T*-TaS$_2$ at doses up to 1 Mrad(SiO$_2$). The threshold voltage changed by less than 2% after irradiation, with persistent self-sustained oscillations observed through the full irradiation sequence. The radiation hardness is attributed to the high intrinsic carrier concentration of 1*T*-TaS$_2$ in both of the phases that lead to threshold switching. These results suggest that charge density wave devices, implemented with thin films of 1*T*-TaS$_2$, are promising for applications in high radiation environments.

*Key Words* — charge density waves, radiation hardness, 1T-TaS$_2$, threshold switching## I. Introduction

Semiconductor devices are susceptible to radiation damage in space and high-energy accelerator environments [1], [2]. Electron–hole pairs generated in the oxide during total-ionizing-dose (TID) irradiation can accumulate in oxide layers and at interfaces in field-effect transistors (FETs) [3]. These can result in shifts of the threshold voltage, increases in static current leakage, errors in bit reading, and, eventually, in complete circuit and/or system failure [1], [2]. Strategies for increasing the radiation resilience of electronic devices typically include improving oxide quality [4], adopting silicon-on-insulator technology [5], and/or using III-V transistors with no oxide layers [6]. Commercial microelectronic devices can often withstand TID exposure to about 1-100 krad(SiO$_2$), and radiation hardened devices typically can withstand doses well above 100 krad(SiO$_2$) [7].

Two-dimensional (2D) materials have shown potential for various electronic applications [8]. Some of us have recently demonstrated a quasi-2D 1*T*-TaS$_2$ device, which operates as a voltage controlled oscillator at room temperature (RT). The physical principal of operation of this oscillator is the transition between two charge density wave (CDW) phases controlled by voltage bias [9], [10]. The transition between the nearly commensurate (NC-CDW) phase and incommensurate (IC-CDW) phase reveals itself as an abrupt change in resistivity of the device channel accompanied by hysteresis. The carrier concentrations in the two CDW states are very high: $10^{21}$ cm$^{-3}$ and $10^{22}$ cm$^{-3}$ for the high resistive NC-CDW and low resistive IC-CDW phases, respectively [11], [12]. Unlike conventional FETs, the 1*T*-TaS$_2$ device is a two-terminal device in which the switching is controlled by the source-drain voltage rather than the gate voltage. No gate oxide is needed for its operation.

In this work, we evaluate the TID response of 1*T*-TaS$_2$ CDW devices by examining the current-voltage (*I-V*) characteristics under X-ray irradiation at doses up to 1 Mrad(SiO$_2$). We find that the threshold voltage, $V_{TH}$, for the abrupt resistance change shifts by only ~2%, the resistance of the CDW states changes by less than ~2 % (low resistive state) and ~6.5 % (high resistive state), and the self-sustained voltage oscillations in this 1*T*-TaS$_2$ oscillator function well through the full irradiation sequence. These results demonstrate the promise of 1*T*-TaS$_2$ CDW devices for use in high radiation environments.

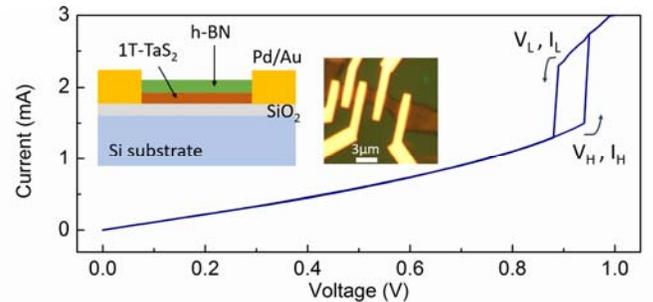

Figure 1: Threshold switching *I-V* characteristics of a 1*T*-TaS$_2$ device at room temperature. The *h*-BN cap on the 1*T*-TaS$_2$ channel is to prevent oxidation during fabrication and air exposure. The left inset shows a schematic diagram of the device structure. The right inset is an optical image of the tested device. The scale bar is 3 µm.

## II. Experimental Results

1*T*-TaS$_2$ devices were fabricated using thin films exfoliated from bulk crystals grown by the chemical vapor transport method [9]. A thin film of 1*T*-TaS$_2$ was placed on a Si

The work of A. A. Balandin and G. Liu was supported, in part, by the National Science Foundation (NSF) through the Emerging Frontiers of Research Initiative (EFRI) 2-DARE project: Novel Switching Phenomena in Atomic MX$_2$ Heterostructures for Multifunctional Applications (NSF EFRI-1433395), the Semiconductor Research Corporation (SRC) and the Defense Advanced Research Project Agency (DARPA) through the Center for Function Accelerated nanoMaterial Engineering (FAME), and by the UC-National Lab Collaborative Research and Training Program (UC-NL CRT). Work at Vanderbilt was partially supported by the Defense Threat Reduction Agency Basic Research Award No. HDTRA1-14-1-0042.

A. A. Balandin and G. Liu are with the Nano-Device Laboratory (NDL) and Phonon Optimized Engineered Materials (POEM) Center, Department of Electrical and Computer Engineering, Bourns College of Engineering, University of California – Riverside, Riverside, California 92521 USA (e-mail: balandin@ece.ucr.edu ; web-site: http://balandingroup.ucr.edu/ ).

E. X. Zhang, C. D. Liang and D. M. Fleetwood are with the Department of Electrical Engineering and Computer Science, Vanderbilt University, Nashville, TN 37235 USA (e-mail: enxia.zhang@vanderbilt.edu).

M. A. Bloodgood and T. T. Salguero are with the Department of Chemistry, University of Georgia, Athens, Georgia 30602 USA.1

substrate with a top layer of 300-nm $SiO_2$. To protect the $1T$-$TaS_2$ thin film channels from oxidization during fabrication and exposure to air, we transferred 10 nm hexagonal boron-nitride ($h$-BN) thin films on top of the channels (see insets to Fig. 1). The electrodes were defined by electron beam lithography. To contact the $1T$-$TaS_2$ thin film underneath the $h$-BN cap, we opened the contact window by dry etching before Pd/Au metal deposition. The thickness of $1T$-$TaS_2$ channel layer ($H$ = 20 nm) has been verified by atomic force microscopy. Details of the device fabrication are reported elsewhere [9].

$1T$-$TaS_2$ devices exhibit a pronounced threshold switching as the voltage exceeds certain threshold values, $V_H$ for the positive scan and $V_L$ for the negative scan. Typical $I$-$V$ characteristics of a tested device at RT are shown in Fig. 1. As the voltage is scanned from 0 V to 1 V, a sudden current jump is clearly observed as the voltage exceeds the threshold value $V_H$ = 0.94 V for this device. At voltages larger than $V_H$, the $I$-$V$ curve has a slope larger than below threshold, indicating a resistance switch from a high resistive state to a low resistive state. As the voltage is scanned back from 1 V to 0 V, the current switches out of the low resistive state to the high resistive state at $V_L$ = 0.89 V. This switching occurs at the electrically driven phase transition between the NC-CDW and IC-CDW phase in $1T$-$TaS_2$ [9]. Since the switching can take place as long as the temperature is below the NC-IC phase transition at 350 K, this operation is stable at RT.

The $1T$-$TaS_2$ device was irradiated with 10-keV X-rays at RT. The $I$-$V$ characteristics of the device were measured after each irradiation step. Fig. 2(a) shows that up to 1 Mrad($SiO_2$), the $I$-$V$ characteristics reveal only a minor change in the threshold voltage. Extracted values of $V_H$ and $V_L$ are plotted as functions of dose in Fig. 2(b). Over the entire exposure range, $V_H$ and $V_L$ change by less than 20 mV, 2.1% and 2.3% of their initial values, respectively. Threshold current values, $I_H$ and $I_L$, corresponding to $V_H$ and $V_L$, change by 2.6 % and 9.6 %, respectively, over the full range of dose. Fig. 2(c) shows that neither the NC-CDW nor the IC-CDW states are sensitive to X-rays. The high resistance value $R_{High}$, extracted from the linear region in the $I$-$V$ curve before switching and the low resistance value $R_{Low}$, extracted from the $I$-$V$ curve after switching, change by 6.5 % and 2.1 %, respectively, up to 1 Mrad($SiO_2$).

We also used the circuit shown in Fig. 3(a) to check whether the $1T$-$TaS_2$ CDW device could function as a self-sustaining oscillator before and after 1 Mrad($SiO_2$) irradiation. The circuit starts to produce an oscillating signal as the voltage across the $1T$-$TaS_2$ device exceeds $V_H$, as shown in Fig. 3(b). The load resistor (1 kΩ used here) provides negative feedback that stabilizes the oscillations. The frequency increases slightly from 1.56 MHz to 1.59 MHz after irradiation, and the amplitude of the oscillations increases from 0.75 V to 0.95 V. The change in performance observed in Fig. 3(b) is due primarily to a change in DC offset during pre- and post-irradiation oscillation measurements, rather than a change in device performance. We also examined Raman spectra of the device channel before and after irradiation to see whether changes in material properties may have occurred, as can happen in graphene as a result of X-ray induced reactions with oxygen, for example [13]. In Fig. 4, the main Raman peaks in the range of 150 cm$^{-1}$ - 450 cm$^{-1}$, corresponding to optical phonon modes, do not show any substantial changes. The bulk and folded optical phonon modes, characteristic for this material [14], [15], remained unchanged before and after irradiation. Hence, we conclude that $1T$-$TaS_2$ CDW devices are capable of performing well in an ionizing radiation environment.

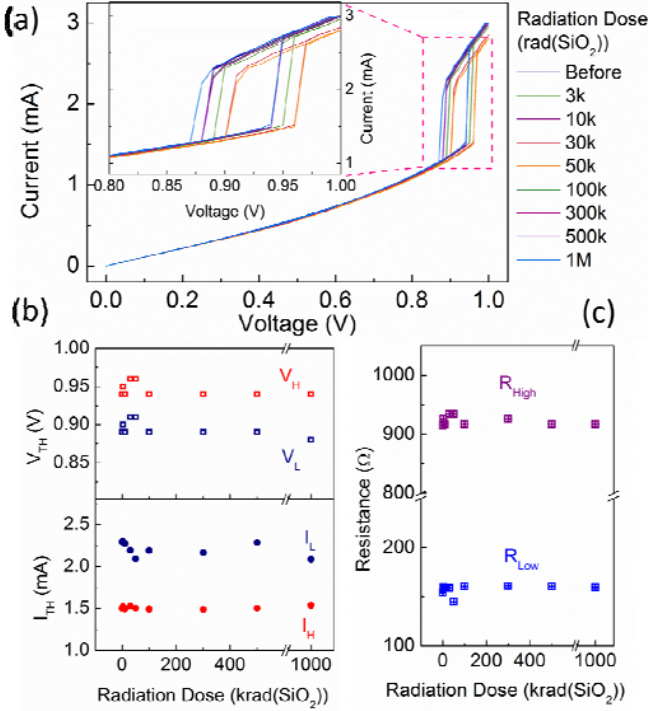

Figure 2: TID response of $1T$-$TaS_2$ devices up to 1 Mrad($SiO_2$). (a) $I$-$V$ curves measured after each X-ray irradiation step. The inset shows a zoomed-in view of the threshold switching region. (b) Threshold voltages, $V_H$ and $V_L$, and threshold currents, $I_H$ and $I_L$, as function of dose. (c) Extracted resistance at the high resistance state and low resistance states as a function of dose.

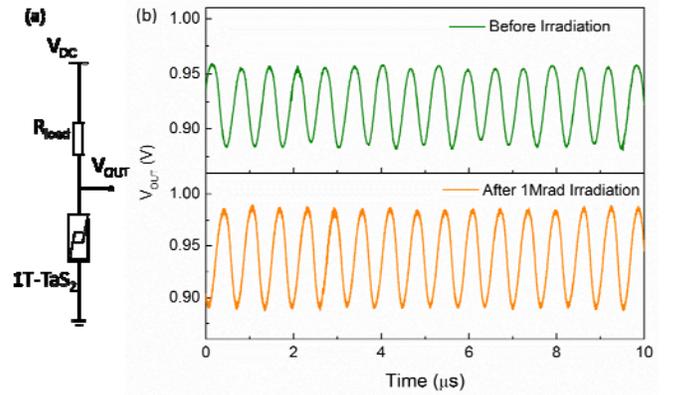

Figure 3: (a) Circuit schematic diagram of a self-sustaining oscillator implemented with one $1T$-$TaS_2$ device and a load resistor. (b) Oscillation waveform before and after 1 Mrad($SiO_2$) X-ray irradiation.

### III. DISCUSSION

We attribute the $1T$-$TaS_2$ device resilience to TID exposure primarily to the two-terminal device design and the very high carrier concentration. The two-terminal device operation is based on the resistance switching controlled by the source-drain voltage, rather than by inducing excess charge via a separate gate voltage through the oxide layer. With no gate or electric field, the charge yield in the h-BN is low [3], and the resulting changes in device characteristics are smaller than those seen in gated graphene-based transistors similarly passivated by h-BN [16]. The other important feature of 1T-TaS$_2$ is the very high carrier concentration in both NC-CDW and IC-CDW states, on the order of $10^{21}$ cm$^{-3}$ and $10^{22}$ cm$^{-3}$, respectively [11], [12]. These values are much higher than those in conventional semiconductor devices; they are closer to those of metals and/or graphene away from the Dirac point [13], [17].

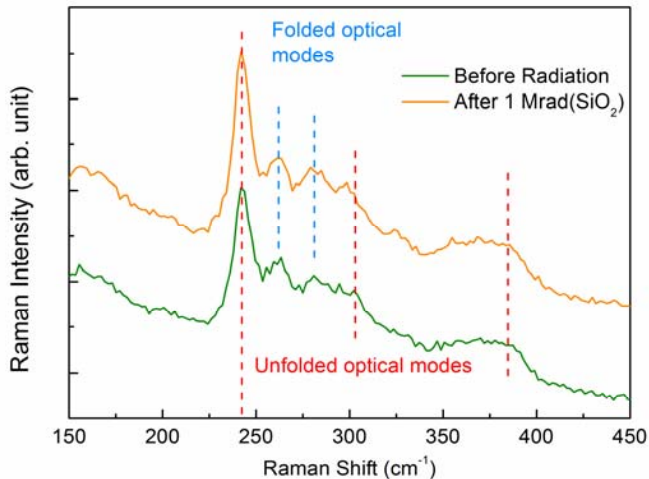

Figure 4: Raman spectra of 1T-TaS$_2$ before and after 1 Mrad(SiO$_2$) X-ray irradiation. The characteristic Raman phonon peaks do not show noticeable changes upon irradiation.

The high carrier concentration in 1T-TaS$_2$ can be attributed, at least in part, to the high quality of the films, which also contributes to reduced radiation-induced material changes. Prior work has shown that exfoliated 1T-TaS$_2$ may contain stacking disorder but not vacancies or other point defects that can act as charge trapping sites [18]. Thus, electron-hole pairs created during irradiation can readily recombine, greatly reducing the charge yield within the 1T-TaS$_2$. The high carrier concentration minimizes the effects of radiation induced charges in the dielectric layers or at interfaces.

Finally, we note that NbO$_x$ and TaO$_x$, in their high resistive states, have rather small carrier concentrations, on the order of $10^{16}$-$10^{17}$ cm$^{-3}$ [19]–[21]. At a similar dose of 0.6 Mrad(NbO$_2$), NbO$_x$ devices have shown 7.1% change in $V_H$, 50% change in $I_H$, and 78.5% change in $R_H$ [22]. In TaO$_x$ devices, just 10 krad(SiO$_2$) irradiation flipped the *OFF* (high resistance) state to *ON* (low resistance) state with a 100-fold decrease in resistance in [21]. On the other hand, HfO$_2$-based resistive memory elements have carrier densities of ~$10^{20}$ cm$^{-3}$ in their conductive state [23], and have been demonstrated to be quite tolerant to TID exposure above 1 Mrad(SiO$_2$) [24], consistent with this interpretation.

IV. CONCLUSIONS

We have investigated the TID response of threshold switching devices made of thin films of the CDW material 1T-TaS$_2$. The electrical characteristics were found to be very robust to X-ray radiation, with only minimal changes observed in switching threshold voltages and currents, and high- and low-state resistances. Moreover, the ability of the device to function as a self-sustaining oscillator was maintained through the full irradiation sequence. The radiation tolerance of these devices is attributed to the two-terminal design and inherently high carrier densities in both the high and low resistant states. These results show that charge density wave devices, implemented with thin films of 1T-TaS$_2$, are promising for applications in high radiation environments.